\documentclass[12pt,nofootinbib]{revtex4}

\usepackage{amsfonts,amssymb}
\usepackage{amsmath}

\usepackage{graphicx}

\def\T11{{T}^{1,1}}
\def\bear{\begin{eqnarray}}
\def\eear{\end{eqnarray}}

\newcommand{\pa}{\partial}
\newcommand{\tr}{{\rm tr}}
\newcommand{\comment}[1]{}
\newcommand{\pasl}{\pa\kern-.55em /}

\newcommand{\ksl}{k\kern-.55em /}

\DeclareFixedFont{\xiiss}{OT1}{cmss}{m}{n}{12}
\DeclareFixedFont{\ixss}{OT1}{cmss}{m}{n}{9}
\DeclareFixedFont{\cmrnine}{OT1}{cmr}{m}{n}{9}
\newcommand{\field}[1]{\mathbb{#1}}

\newcommand{\BC}{{\field C}}

\newcommand{\BZ}{{\field Z}}

\newcommand{\CCs}{\hbox{\ixss C\kern-.4emI}}
\newcommand{\ZZs}{\hbox{\ixss Z\kern-.4emZ}}

\usepackage{amsmath}
\usepackage{graphicx}

\newcommand{\expect}[1]{\langle #1\rangle}

\newcommand{\der}[2]{\frac{\partial #1}{\partial #2}}
\begin{document}
\title{Counting conifolds and Dijkgraaf-Vafa matrix models for three matrices }
\author{David Berenstein\\ 
Department of Physics, UCSB, Santa Barbara, CA 93106\\
Email: \email{dberens@physics.ucsb.edu}}
\author{Samuel Pinansky\\
Department of Physics, UCSB, Santa Barbara, CA 93106\\
Email: \email{samuelp@physics.ucsb.edu}}
\begin{abstract}
We study superpotential perturbations of $q$ deformed ${\cal N}=4$ Yang-Mills for $q$ a root of unity. This is a special case whose geometry is associated to an orbifold 
with three lines of codimension two singularities meeting at the origin. We perform
field theory perturbations 
 that leave only co-dimension three singularities of conifold type in the geometry. We show that there are two ``fractional brane'' solutions of the F-term equations for each singularity in the deformed geometry, and that the number of complex deformations of that geometry also matches the number of singularities. This proves that for this case there are no local or non-local obstructions to deformation.  We also show that the associated Dijkgraaf-Vafa matrix model has a solvable sector, and that the loop equations in this sector encode the full  deformed geometry of the theory.
 \end{abstract}

%%\pacs{11.25.Tq, 11.15.Pg}

\maketitle

\tableofcontents
%%\keywords{Matrix models,  Supersymmetric gauge theories}
%\preprint{hep-th/0602294}

\section{Introduction}

Supersymmetric field theories are a very attractive area of modern theoretical physics not only because of their possible application to physics beyond the standard model, but also
because they have features that can be solved exactly non-perturbatively \cite{Shol}. 
The program of geometric engineering produces these type of field theories by placing 
compact D-brane configurations on non-compact Calabi-Yau manifolds. One of the main features of using non-compact CY manifolds is that four dimensional gravity decouples, and in the low energy limit we are left with an ordinary supersymmetric field theory. 
The advantage of such a formulation is that many aspects of the field theory can be understood geometrically. For example, the holomorphic structure of the Calabi-Yau manifold might be deformed at singularities, and this can be associated to confinement \cite{KS,GV,CV} 

Having this geometric information is important in terms of the AdS/CFT correspondence if we want to understand gravity duals to certain field theories, and it is also useful for model building if we want to have a theory on a CY throat that describes the standard model of particle 
physics (or some similar theory) via a D-brane configuration. 
In practice it is very hard to compute the low energy field theory on the D-brane because we need to have full knowledge of the superpotential, and this can involve calculating arbitrarily complicated $n$ point functions on the D-brane. This can be overcome if we have a large amount of symmetry that forbids most terms (this is what happens in practice for CFT's associated to toric CY
singularities, see for example \cite{BHK} ).

We can also try to work in the inverse direction. Start with some field theory and try to construct a dual geometry that realizes it.  Some progress in this direction has been made 
in \cite{BL, Brev}, by turning the problem into calculating certain properties and features of an algebra associated to the field theory. 
 However, there is no algorithm to systematically carry out the procedure outlined there (except when there is a lot of symmetry), even at the classical level, and one has to work with specific examples.

One of the big discoveries of the geometric engineering program was the realization by Dijkgraaf and Vafa that one can calculate an effective superpotential for gaugino condensation by studying the planar diagrams of an associated matrix model derived from the superpotential \cite{DV,DV2,DV3}. 
This has been derived exactly in field theory in the works \cite{DGLVZ,CDSW}.
In many examples, one only has to worry about a one-matrix model, 
and one can relate the spectral curve of the matrix model to the (deformed) geometry of the original Calabi-Yau space. This usually works best if one can understand the  model as a deformation of an ${\cal N}=2 $ theory, where one also can solve the model using Seiberg-Witten techniques \cite{SW}. 

In most cases however, one has to try and understand the problem in terms of a more generic looking matrix model. In this case it is not obvious that there is a spectral curve, or how the geometry of the CY manifold is related to the matrix model. A proposal for what one is  supposed to do in this case has been outlined in \cite{Bq}, by studying the quantum moduli space of a brane in the bulk, but to date there is no proof that this always works. One can also encode in some examples  the CY complex structure geometry by looking at the loop equations associated to the matrix model \cite{Bhol}, see also \cite{Fer,Bon} for other examples.

The purpose of this paper is to explore this issue for a particular class of models. The strategy that one can adopt is to begin in a configuration where one knows the field theory very precisely. Then one has a dictionary between certain superpotential deformations and certain deformations of the geometry. This is usually simple to understand in the case where one has 
(locally) an ADE orbifold singularity (these are codimension two singularities in the CY geometry) and one can identify how the closed string moduli that resolve the singularity couple to a D-brane.
In this case, one can follow the deformation at the D-brane level (field theory)  as well and one can try to reconcile both geometries. In our examples we begin with the $C^3/\BZ_n\times \BZ_n$ orbifold in the presence of discrete torsion and follow that general idea in the example.
A generic deformation will resolve the codimension two singularities and might leave behind some codimension three singularities in the geometry. Most of the time, these are just
conifold singularities (these are all the singularities we find in the generic case), and we understand how these get deformed locally by placing fractional branes at the singularity \cite{KS} \footnote{These brane configurations can also produce duality cascades in the field theory for the low energy dynamics of branes at the singularity.  It also seems to be the case that the field theory associated to conifolds is always in the same universality class \cite{Bcon}. Recently, the understanding of the full moduli space of vacua of these cascading theories has been improved substantially \cite{DKS} }

 We expect this behavior to be universal in the presence of fractional branes at the singularity and to have one deformation parameter per conifold singularity. However, there always exist the possibility that some of these deformations are obstructed (perhaps non-locally), or that the naive geometric intuition does not match the field theory analysis. Indeed, examples have been found where there is an inconsistency of the chiral ring once gaugino condensates are taken into account \cite{BHOP,dP1}, and one finds that supersymmetry is broken or there is runaway behavior \cite{IS}.
 Morever, even if we understand the local behavior at each singularity, we might have trouble understanding the global 
 deformed geometry in detail.  This is especially hard in the case of non-complete intersection varieties, although recent progress has been made for the special case of toric geometries \cite{ME}.
 In this paper we  show that for the example we study there are no such obstructions and that the field theory analysis reproduces the geometric analysis precisely. 

The model we work with is built by starting with the $q$ deformed ${\cal N}=4 $ SYM theory, which is associated to the $\BC^3/{\BZ_n\times \BZ_n}$ orbifold with discrete torsion \cite{D,BL2}. The theory has the same matter content of ${\cal N}=4 $ SYM, but a different superpotential
\begin{equation}
W = \tr(XYZ-qXYZ)
\end{equation}
with $q^n=1$. This model has been studied extensively in the past, in various contexts, see for example \cite{LS,BJL,DHK,BC}, and more recently 
the interest in the model for general $q$ has been sparked by the dual supergravity solution found by Maldacena and Lunin \cite{LM}.
All the models for different roots of unity give rise in the Dijkgraaf-Vafa program to three-matrix models which on the face look very similar to each other. The associated geometries end up being fairly different, although one can still study them in a unified context.

Our reasons for studying this particular model are various. The theory has a unique gauge group with a simple matter content. The corresponding matrix model has three matrices and looks simple as well. On the other hand the geometric interpretation is relatively complicated and leads to a very useful laboratory to study 
complex structure deformation problems.

The idea is now to deform the field theory superpotential 
given above by twisted sector deformations. These deformations are generated by
\begin{equation}
\tr(X^k), \tr(Y^k), \tr(Z^k)
\end{equation}
for $k$ not a multiple of $n$ \cite{BL2}. These are non-zero elements of the chiral ring whose support 
is concentrated at the codimension two singularities of the $\BC^3/\BZ_n\times \BZ_n$
orbifold space. Because these are elements of the chiral ring with support on the singularities, these are associated to complex structure deformation moduli in string theory. This is what one obtains
from an analysis within the framework of the AdS/CFT correspondence \cite{M,Guk,BL2}.

However, we have to deal with the problem that the codimension two singularities are non-compact and there are an infinite number of possibilities, some of which might deform the geometry severely. This same problem is present in the work on deformations of $N=2$ theories \cite{CV}. The idea is to think of these deformations we are performing as being  marginal or relevant. 
This depends on a notion of $R$-charge, that might differ from the naive one that one deduces directly from the $q$ deformed ${\cal N}=4$ theory, as if one had a different UV fixed point. It must also be the case that since we want the theory to be dominated by the ${\cal N}=4$ $q$ deformed superpotential, that this represents a marginal deformation. Obviously we have a $U(1)^3$ symmetry of the $q$ deformed potential.
With this symmetry we can assign different R-charges to $X,Y,Z$ so that they are all positive
and given by $\gamma_1, \gamma_2, \gamma_3$ respectively, and so that the $R$ symmetry of the $q$ deformed superpotential is exactly two. This type of modification is similar to what is done in \cite{BSoo} for non-anticommutative power counting. Here, it modifies the notion of relevant operators for superpotential couplings, which are the ones that control the holomorphic data of the theory. This in the end tuns out to be  a convenient tool to keep track of holomorphy without doing a complicated analysis of charges.

With this prescription, we are only allowed to add monomials $\tr (X^k)$ such that $k\gamma_1\leq 2$
$\tr(Y^l)$ so that $l\gamma_2\leq 2$ and $\tr(Z^m)$ so that $m\gamma_3\leq 2$.

Indeed, we can choose $\gamma_3$ arbitrarily close to one, so we can deform both in the $X$
and $Y$ singularity as much as we want to, and then the only extra term that depends on $Z$ is exactly $\tr(Z)$. For the sake of simplicity, we will take the coupling associated to $\tr(Z)$ to be exactly zero. In principle one can repeat the analysis we have done here for this slightly more general case. Part of this analysis at the classical level for a particular case has been done in \cite{DF} however, we have not yet found a closed form expression of the geometry for  the full theory in the more general case.

The general geometry that we get without gaugino condensations is described by the following simple hypersurface equation
\begin{equation}
uvw=Q(t,u)+Q'(t,v)-t^n
\end{equation}
where the orbifold with discrete torsion corresponds to the case $uvw=t^n$. $Q$ and $Q'$ 
are polynomials with leading term $t^n$, in fact
it will turn out that the polynomials we find can be written in factorized form
\begin{align}
Q(t,u)&=\prod_{i=0}^{n-1}(t-f(q^i u^{1/n}))\\
Q'(t,v)&=\prod_{i=0}^{n-1}(t-g(q^i v^{1/n}))
\end{align}
where on the right hand side we have to choose some n-th root of $u,v$ and where $f(0)=g(0)=0$. It is easy to see that
it does not matter which root we choose, we get the same end result, and that $Q$ and $Q'$
are indeed polynomials of $u,v$.
It is clear that $Q$ has singularities when different roots of $Q$ at $u$ fixed coincide. The same is true
for $Q'$. These give rise to singularities of the full geometry when either $u$ or $v$ are zero. These are generically conifold singularities.
It is easy to convince oneself that these are the only singularities in the above geometry. The derivative with respect to $w$ forces either $u$ or $v$ to be zero. After this is used, one is left with either $Q$ or $Q'$ only. The derivative with respect to the variable that is turned to zero is used in turn to determine the value of $w$.

The main issue we will be concerned with is to count all the conifold points of this geometry. 
We will show that for each of them there is a log-normalizable deformation of the geometry in a suitable sense, and we will show that for each of these singularities one also finds a pair of fractional branes
that can get trapped at the singularity. We furthermore show that if we take into account the loop equations of the associated matrix model, that we can recover the full geometry, and that there are no 
field theory obstructions to deformation. 

This fits well with the conjectures in \cite{BHOP}, where it is stated that 
only in the case of geometric obstructed deformations we should get an inconsistency of the quantum deformed chiral ring of the theory.

In a sense, our results are exactly as they should be and provide a technically involved consistency check of the Dijkgraaf-Vafa ideas, as well as an interesting solution to a complicated multi-matrix model. Presumably these techniques can be extended to other setups with multiple gauge groups as well.  A very interesting case to analyze is a perturbation of 
the $\BC^3/\BZ_2\times \BZ_2$ singularity without discrete torsion. In that case it has been observed that fractional branes at the origin trigger runaway behavior for fractional branes on the codimension two singularities \cite{Brun}. Maybe in this case one would obtain a more involved 
obstruction to deformation.

The paper is organized as follows:  In section 2 we will analyze the deformed ${\cal N}=4$ theory by solving the F-term constraints to find solutions corresponding to branes in the bulk as well as fractional brane solutions at singularities, and find the hypersurface equation defining the deformed geometry.  We will also count the number of relevant quantum deformations, and show that it is equal to the number of singularities of the geometry.  In section 3 we will analyze the theory by finding the quantum relations of the associated matrix model, and show how the loop equations encode the complex structure of the CY geometry.

\begin{section}{Deformed ${\cal N}=4$ Superpotential}

The unique ${\cal N}=4$ supersymmetric gauge theory in four dimensions has three adjoint fields $X,Y,Z$ in the gauge group $G$ and a superpotential
\begin{equation}
\tr(XYZ-XZY).
\end{equation}
For our purposes $G$ will be $SU(N)$. We are interested in studying the structure  of 
closely related field theories with the same field content, but with a different 
superpotential. Some of these are very interesting models in geometric engineering because they correspond to the field theories associated to D-branes in singular geometries. Some of these are orbifolds with discrete torsion $\BC^3/\BZ_n\times \BZ_n$ \cite{BJL,D}. For this case 
the superpotential is:
\begin{equation}
W_0=\tr(XYZ-qXZY)
\end{equation}
where $q^n=1$ is a root of unity. The geometry is given by the hypersurface in $\BC^4$
defined by
\begin{equation}
u v w = t^n \label{eq:orb}
\end{equation}
and has three lines of $A_n$ singularities (along the $X,Y,$ and $Z$ axis) of codimension two which intersect at the origin. At a generic point of the singular
locus the geometry looks locally like a $\BC^2/\BZ_n\times \BC$ quotient and the low energy effective field theory of a brane at that locus has an ${\cal N}=2$ supersymmetry. Globally this is not true, but one can analyze the theory in the same spirit of Seiberg-Witten. This task has been 
performed in a series of papers, mostly in the work of Dorey et al \cite{Detal}.
The relation to AdS/CFT has also been extensively studied, mostly because these theories belong to  a one parameter family of marginal deformations of ${\cal N}=4$ SYM \cite{LS} and as such one could expect that for sufficiently small deformations one can understand the dual $AdS$ space at the level of a supergravity analysis \cite{BL, BJL,Ket,LM}.

Perhaps the biggest surprise in  the study of this class of deformations is that the field theory is so  rich, in spite of looking deceptively close to ${\cal N}=4 $ SYM.
We wish to further deform the geometry to break these lines of singularities down to points, i.e. codimension three singularities. This is accomplished by turning on additional 
superpotential term couplings. 

We want to argue that in some sense these are relevant deformations of the superpotential, and that the superpotential $W_0$ dominates the geometry. This can only be argued if the 
superpotential deformation satisfies some additional constraints; the ones that are  relevant by power counting have been analyzed in \cite{BJL, D,DF} . There is also the Polchinski-Strassler setup \cite{PS}, that does not fall into the category of deformations that we allow for $q=1$.

Here we will give a definition of what relevant terms mean in a holomorphic sense, from the point of view of deformation theory.
The idea is simple, the hypersurface equation $uvw=t^n$ has a $U(1)^3$ symmetry 
under independent rescalings of $u, v, w$ ($t$ gets rescaled by each of these transformations). We can choose from this set a generalized degree for $u,v, w$ $n r_1,n r_2, n r_3$ such that they all have positive degree, and such that the degree 
of $t$ is one. In this way $r_1+r_2+r_3=1$. A complex structure deformation of the equation will be given by a polynomial $f$ in the variables $u,v, w, t$, so that the hypersurface equation reads $uvw= t^n+f$. The deformation associated to $f$ will be relevant if all the monomials of $f$ have strictly smaller degree than $n$. At the level of the field theory,  $X, Y, Z$ will have charge $r_1, r_2, r_3$ with respect to this $R$-like symmetry, and the deformation of the superpotential will be a polynomial in $X, Y, Z$ of degree less than one. The standard choice is
$r_1=r_2=r_3=1/3$. We will consider a different choice where $r_3$ is arbitrarily close to $1$. In this sense, we can allow any deformation of the superpotential which is a polynomial in $X, Y$, but which does not depend on $Z$. We will also not consider a linear term in $Z$ because this simplifies the analysis.

\subsection{F-term equations}
We start from the most general form for a deformed superpotential with three fields $X,Y,Z$ in the adjoint:
\begin{equation}
W=\tr(XYZ-qXZY+V(X)+U(Y))
\end{equation}
where
\begin{align}
q^n&=1\\
V(x)&=\sum_{i\neq 0\mod n}\frac{a_i}{i}x^i\\
U(y)&=\sum_{i\neq 0\mod n}\frac{b_i}{i}y^i
\end{align}
The restrictions on $V$ and $U$ guarantee that this superpotential has a geometric brane realization. This is because the generic complex deformation we are turning on should correspond to twisted sector moduli localized at the codimension two singularities (these are the ones that resolve the singularities). The terms $X^kY^m$ have $\BZ_n\times\BZ_n $ charge $(k,m)\mod(n)$ under the quantum symmetry of the orbifold. 
Although the form above seems non-generic, note that any  cross terms (with non-trivial twisted sector charge) of the form $\tr(X^kY^m)$ are proportional to a polynomial in $X$ or $Y$ plus terms
proportional to the F-term equations of motion for the field $Z$. These can be absorbed in a redefinition of $Z$ by a linear shift.
These conditions also make it so that the F-term equations are solvable in terms of $n\times n$ matrices (this corresponds to a brane in the bulk  in the underformed case), and the moduli space of the brane in the bulk is exactly the deformed geometry. We will now derive this from studying the $F$ term constraints.
We have three matrix equations derived from the F-term constraints:
\begin{align}
XY-qYX&=0\label{ft1}\\
ZX-qXZ&=-\sum_{i\neq 0\mod n}b_i Y^{i-1}\label{ft2}\\
YZ-qZY&=-\sum_{i\neq 0\mod n}a_i X^{i-1}\label{ft3}
\end{align}
and we want to find all  solutions of these matrix equations, up to equivalence. This is the problem of solving the classical moduli space of vacua of the theory.

 In general,
if we have two different solutions, we can take direct sums, and they also solve the 
equations. The most interesting solutions are those that are given by irreducible solutions
(those that can not be written as direct sums of smaller solutions. These are the ones that we associate to individual branes following \cite{BL,Brev}

\begin{subsection}{Bulk Brane Solutions}
We first find the solutions to the F-term equations of dimensionality $n$, since those representations correspond
to branes in the bulk.  We begin by solving equation \eqref{ft1}.  We can choose $X$ to be diagonal, and then the general solution is of the form 
\begin{equation}
X=xP\quad Y=yQ
\end{equation}
where $P$ and $Q$ have the properties $P^n=Q^n=1$, $PQ=qQP$, and $x$ and $y$ are free complex parameters different from $0$. 
When any of these two go to zero, one has to be more careful in describing the geometry, because either $X^n$ or $Y^n$ become zero. This means that all eigenvalues of $X$ or $Y$ go to zero, but the matrices may still be upper or lower triangular, and this is necessary to be able to satisfy the equations (\ref{ft2}) and (\ref{ft3})

With $P$ chosen diagonal, the matrices are uniquely:
\begin{equation}
P=\left(\begin{array}{ccccc}
1&&&&\\
&q&&&\\
&&q^2&&\\
&&&\ddots&\\
&&&&q^{n-1}
\end{array}\right),
Q=\left(\begin{array}{ccccc}
0&0&\dots&0&1\\
1&0&\dots&0&0\\
0&1&\dots&0&0\\
\vdots&\vdots&\ddots&\vdots&\vdots\\
0&0&0&1&0
\end{array}\right)
\end{equation}
the so called {\em{clock}} and {\em{shift}} matrices.  Now we insert these into the equation \eqref{ft2}:
\begin{align}
ZX-qXZ=-\sum_{i\neq 0\mod n}b_iy^{i-1}Q^{i-1}
\end{align}
It is a well known fact that the $P$ and $Q$ matrices span the $n\times n$ complex matrices as a vector space, so we can write $Z$ in general as $Z=\sum_{i,j=0}^{n-1}c_{ij}Q^iP^j$.
Inserting this into the equation allows us to solve it:
\begin{align}
\sum_{i,j=0}^{n-1}c_{ij}xQ^iP^{j+1}-\sum_{i,j=0}^{n-1}c_{ij}xqPQ^iP^j&=-\sum_{i\neq 0\mod n}b_i y^{i-1}Q^{i-1}\\
\sum_{i,j=0}^{n-1}c_{ij}xQ^iP^{j+1}(1-q^{i+1})&=-\sum_{i\neq 0\mod n}b_i y^{i-1}Q^{i-1}
\end{align}
Considering the linear independence of the matrices, we can equate terms with equal powers of $Q$ and $P$, giving for
$i+1\neq 0\mod n$:
\begin{align}
\label{linind}c_{i-1,n-1}x(1-q^i)&=-\sum_{j=i\mod n}b_jy^{j-1}\\
c_{i-1,n-1}&=-\sum_{j=i\mod n}x^{-1}\frac{b_{j}y^{j-1}}{1-q^i}
\end{align}
For the other case, $i+1=0\mod n$:
\begin{align}
c_{n-1,j}xP^j(1-q^{n})&=0\\
0&=0
\end{align}
is satisfied for all $c_{n-1,j}$.  Let us now relabel our sum and call $c_{i-1,n-1}=z_i'$ ($i$ runs from $1$ to $n-1$).
We can rewrite  $Z$  to make it a bit clearer:
\begin{equation}
Z=\sum_{i=1}^{n-1}z_i' Q^{i-1}P^{n-1}+\sum_{i=0}^n c_{n-1,i}Q^{n-1}P^{i}
\end{equation}
Now insert this form into equation \eqref{ft3} 
\begin{align}
YZ-qZY&=-\sum_{i\neq 0\mod n}a_i X^{i-1}\\
\sum_{i=1}^{n-1}z_i'yQ^iP^{n-1}(1-q^n)+\sum_{i=0}^nc_{n-1,i}yqP^i(1-q^i)&=-\sum_{i\neq 0\mod n}a_i x^{i-1}P^{i-1}\\
\sum_{i=0}^nc_{n-1,i}yP^i(1-q^{i+1})&=-\sum_{i\neq 0\mod n}a_i x^{i-1}P^{i-1}
\end{align}
Exactly as in equation \eqref{linind} above we equate terms with the same powers of $P$ to get:
\begin{equation}
c_{n-1,i-1}=-\sum_{j=i\mod n}y^{-1}\frac{a_{j}x^{j-1}}{1-q^i}
\end{equation}
with $i=1..n-1$, and that $c_{n-1,n-1}$ is unrestricted.  We now make the definitions: $c_{n-1,i-1}\equiv z_i$ and
that $c_{n-1,n-1}\equiv z$.  We can now write down the most general $n$ dimensional solution to the F-term equations:
\begin{align}
X&=xP\\
Y&=yQ\\
Z&=zQ^{n-1}P^{n-1}+\sum_{i=1}^{n-1}\left(z_i Q^{n-1}P^{i-1}+z'_iQ^{i-1}P^{n-1}\right)
\end{align}
where the $z_i, z'_i$ are given by
\begin{align}
z_i&=-\sum_{j=i\mod n}y^{-1}\frac{a_{j}x^{j-1}}{1-q^i}\\
z'_i&=-\sum_{j=i\mod n}x^{-1}\frac{b_{j}y^{j-1}}{1-q^i}
\end{align}
This gives us  a family of solutions parameterized by three complex numbers $x,y,z$, and therefore correspond to branes free to propagate in the bulk (which is a three dimensional CY geometry). However, the parameters $x,y,z$ are not gauge invariant, as we have implicitly made gauge transformations by selecting $X$ diagonal. One can show that $x$ transforms as $x\to qx$ by conjugating the solution with $Q$. This also transforms $z$ and the $z_i,z_i'$, so the parameters $x,y,z$ give a branched covering of the moduli space.

We want to find the moduli space more explicitly as a hypersurface equation in $\BC^4$, where we can make contact with the original, undeformed geometry.

To find the proper equation, we need to find the gauge invariant observables of the configurations, and this can be translated into the center of the algebra defined by the three equations (the center is the set of all polynomials in $X,Y, Z$ which when evaluated in the solutions we found give matrices proportional to the identity.) This has been advocated as the correct description of D-branes at singularities in \cite{BL}.

Note that $X^n=x^n=\det X\equiv u$ and $Y^n=y^n=\det Y\equiv v$ are obviously in the center.  It is also clear that $Z^n$ is
not in the center (which can be readily verified from the algebra).  However, one can find that $\det Z\equiv w$ is, by construction, in the center.  (A note of caution: $w$ is not obviously a polynomial of $X,Y,Z$, but after some algebra we will find that it is  ). A final variable in the
center is given by $XYZ+f(X)+g(qY)\equiv t$, where $f$ and $g$ are polynomials.  We can solve for them easily given the above solution:
\begin{align}
XYZ&=xyz-\sum_{i=1}^{n-1}\left(xyz_iP^i+xyz_iq^iQ^i\right)\\
&=xyz-\sum_{i\neq 0\mod n}\frac{a_i}{1-q^i}X^i-\sum_{i\neq 0\mod n}q^i\frac{b_i}{1-q^i}Y^i
\end{align}
so
\begin{align}
\label{deff}f(X)&=\sum_{i\neq 0\mod n}\frac{a_i}{1-q^i}X^i\\
\label{defg}g(Y)&=\sum_{i\neq 0\mod n}\frac{b_i}{1-q^i}Y^i
\end{align}
Note that $f$ and $g$ are polynomials in $X$ and $Y$ of the same order as $V$ and $U$ respectively.  Also, we have that
$f(x)=xy\sum z_i$ and $g(y)=xy\sum z'_i$.  
\par
An aside before we derive the hypersurface equation:  How do we know that $\det Z$ is a polynomial in the algebra?  
We note that we can solve for the $z_i$'s in terms of $f(q^ix)$, specifically
\begin{equation}
-\frac{1}{nxy}\sum_j q^{-ij}f(q^j x)=z_i
\end{equation}
and likewise for the $z'_j$'s.  We can use the characteristic equation for the $Z$ matrix, which in general is of the form
\begin{equation}
\pm \lambda^n+\tr(Z)\lambda^{n-1}+\cdots+\det Z=0
\end{equation}
where the omited terms are polynomial combinations of higher power traces.  Note first that $\tr(Z)=0$, so there
is no $n-1$ power term.  Plugging in the matrix $Z$ itself into the characteristic equation (it must be a solution
to its own characteristic equation by the Cayley-Hamilton theorem), we see that 
\begin{equation}
\det Z=\pm Z^n-{\rm polynomial\;in\;}Z{\rm \;of\;degree\;} n-2
\end{equation}
where the coefficients of the polynomial are combinations of the coefficients of $Z$ we found above.  As we showed,  the coefficients of
$Z$ are simply the $z_i$'s (and some $q$'s), which can be written in terms of polynomials of the $X$ or $Y$ matrices (we must allow for dividing by $x,y$, but these are non-zero). It turns out that these terms in the denominators cancel.
\par
We can continue on to obtain the full hypersurface equation.  Using the expression for $t$, we simply take the determinant:
\begin{align}
XYZ&=t-f(X)-g(qY)\\
uvw&=\det(t-f(X)-g(qY))
\end{align}
This is relatively easy to analyze, since the $t-f(X)$ is diagonal and the $-g(qY)$ is strictly off diagonal.  This provides
a factorization:
\begin{equation}
uvw=\prod_{i=0}^{n-1}(t-f(q^ix))+{\rm terms\;with\;}g(y){\rm 's}
\end{equation}
We can go further.  From the symmetry between $X$ and $Y$ present in the superpotential, we could have equally
likely chosen $Y=yP$ and $X=xQ$, so there must be symmetrical terms for each $u$ and $v$.  This gives:
\begin{equation}
uvw=\prod_{i=0}^{n-1}(t-f(q^ix))+\prod_{i=0}^{n-1}(t-g(q^iy))-t^n+{\rm cross\;terms\;with\;at\;least\;one\;} f{\rm \; and \;}g
\end{equation}

From the original determinant expression, we also see that the RHS is invariant under $x\to qx$ and $y\to qy$
($x\to qx$ implies $f(X)\to Q^iX^iQ^{-i}$ for each term in $f(X)$ which is invariant under the determinant, and $y\to qy$ implies likewise with
$P$ instead).  From this, we know that the right side can only be a function of the invariants under these transformations, namely
$u$ and $v$.  Then from our further factoring, we can say that
\begin{equation}
uvw=Q(t,u)+Q'(t,v)-t^n+uv R(t,u,v)
\end{equation}
where $R$ has degree less than $n-1$ in all it's variables, and
\begin{align}
Q(t,u)&=\prod_{i=0}^{n-1}(t-f(q^ix))\\
Q'(t,v)&=\prod_{i=0}^{n-1}(t-g(q^iy)
\end{align}
We can now do a change of variables, shifting
$w\to w+R(t,u,v)$, giving us our final hypersurface equation
\begin{equation}
uvw=Q(t,u)+Q'(t,v)-t^n
\end{equation}
which is surprisingly simple to write and analyze. When either $g$ or $f$ are zero it reduces to 
known geometries (see for example, \cite{HK}), and turning on both gives the simplest geometry that one could possibly construct that captures both limits correctly. This is because anything more complicated that could appear on the right hand side could be shifted away by a coordinate redefinition of $w$. Notice that
the only hard part of the analysis is to actually find $w$ as a polynomial in $X,Y, Z$, which we did implicitly. The most important thing is that in spite of beginning with some complicated looking solutions of the $F$-terms, the final form of the geometry is very simple.

\end{subsection}
\begin{subsection}{Fractional Brane Solutions}

The solutions we found above correspond to the generic bulk brane. It is interesting to find out 
if there are some cases where we can find solutions of smaller rank than $n$. In the case of orbifolds these solutions correspond to the existence of fractional branes at the singular locus.
We will call these solutions the fractional brane solutions, and we will give them the standard interpretation, namely, as branes stuck at singularities of the geometry we found. Also, since it is easy to write down $u,v,t$ in the geometry as polynomials of $X,Y,Z$, and to use that information to calculate $w$, we will calculate these coordinates for the fractional brane solutions explicitly.

Let us enumerate solutions with dimension $p<n$. If we begin by taking $X$ to be diagonal, then equation \eqref{ft1} gives:
\begin{align}
x_iy_{ij}&=qx_jy_{ij}
\end{align}
with no sum implied.  We want solutions where both $X$ and $Y$ are non-zero, which gives $y_{ii}=0$.  We can also satisfy 
$p-1$ of the rest of the equations by choices of the $x_i$'s.  We choose without loss of generality that $qx_i=x_{i+1}$.  Therefore only $y_{i+1,i}$ are non-zero, with the
rest of the entries in $Y$ zero.  Relabeling $x_1=x$ and choosing a gauge where all the $y_{i+1,i}$'s are $1$, gives the
following most general form for $X$ and $Y$ solving the first F-term equation \eqref{ft1}:
\begin{align}
X=x\left(\begin{array}{cccc}
1&0&\dots&0\\
0&q&\dots&0\\
\vdots&\vdots&\ddots&\vdots\\
0&0&\dots&q^{p-1}
\end{array}\right)\quad 
Y=\left(\begin{array}{cccc}
0&0&\dots&0\\
1&0&\dots&0\\
\vdots&\ddots&\ddots&\vdots\\
0&\dots&1&0
\end{array}\right)
\end{align}
\par
To solve equation \eqref{ft2}, we leave $Z$ an arbitrary $p$ by $p$ complex matrix, and plug into equation \eqref{ft2} our
known form for $X$ and $Y$:
\begin{align}
ZX-qXZ&=-\sum_{i\neq 0 \mod n}b_iY^{i-1}\\
x\left(\begin{array}{cccc}
z_{11}(1-q)&\dots&z_{1j}(q^{j-1}-q)&\dots\\
\vdots&\ddots&\dots&\dots\\
z_{i1}(1-q^i)&\dots&z_{ij}(q^{j-1}-q^i)&\dots\\
\vdots&\vdots&\ddots&\dots
\end{array}\right)&=-\left(\begin{array}{ccc}
b_1&0&\dots\\
b_2&b_1&\dots\\
b_3&b_2&\ddots\\
\vdots&\ddots&\ddots
\end{array}\right)
\end{align}
Note that the sum truncates since $Y^p=0$.  These equations are solved by sets of $z$'s such that
\begin{align}
z_{ij}&=qz_{i+1,j+1}\\
z_{i1}&=-\frac{b_i}{x(1-q^i)}\\
z_{ij}&=0\quad {\rm for\;}i+1<j
\end{align}
with the $p-1$ elements $z_{i,i+1}$ still unrestricted.  The other elements are uniquely fixed by the above recursion relation.
\par
To solve equation \eqref{ft3}, we plug in our forms giving:
\begin{align}
YZ-qZY&=-\sum_{i\neq 0 \mod n}a_i X^{i-1}\\
\left(\begin{array}{cccc}
-qz_{12}&-qz_{13}&\dots&0\\
z_{11}-qz_{22}&z_{12}-qz_{23}&\dots&z_{1p}\\
\vdots&\vdots&\vdots&\ddots
\end{array}\right)&=-
\left(\begin{array}{ccc}
V_1'(x)&0&\dots\\
0&V_1'(qx)&\dots\\
\vdots&\vdots&\ddots
\end{array}\right)\label{fb1}
\end{align}
Analyzing \eqref{fb1}, we see that given the relationships between the $z_{ij}$ required to satisfy \eqref{ft2}, 
all but the diagonal equations are automatically satisfied.  Using the relation $xV'(x)=f(x)-f(qx)$, with $f$ defined
as above, we get a system of $p$ equations:
\begin{align}
-xqz_{12}&=-f(x)+f(qx)\\
qx(z_{12}-qz_{23})&=-f(qx)+f(q^2x)\\
\vdots&=\vdots\\
q^{p-2}x(z_{p-2,p-1}-qz_{p-1,p})&=-f(q^{p-2}x)+f(q^{p-1}x)\\
q^{p-1}xz_{p-1,p}&=-f(q^{p-1}x)+f(q^px)
\end{align}
We have $p$ equations with $p-1$ variables, so the equations must not be linearly independent to be solvable.  Adding
the equations together collapses nicely, giving:
\begin{equation}
f(x)=f(q^px)
\end{equation}
Therefore only for values of $x$ satisfying this equation will there exist a fractional brane solution with dimension $p<n$.
\par
We could have also taken $Y$ diagonal in the first step, interchanging the role of $X$ and $Y$.  An identical derivation
to the one above gives that solutions exist only at points such that $g(y)=g(q^py)$, where $y$ is the element $y_{11}$ of $Y$.
Thus we have two different sets of fractional brane solutions for each $p$, of number approximately equal to the degree of the polynomial $f(x)-f(q^px)$ plus the
degree of $g(y)-g(q^py)$.  For each of these either $X^n$ or $Y^n$ are zero, this is $u$ or $v$ vanish at that point. 

\end{subsection}
\begin{subsection}{Singularities of the hypersurface equation}
We derived the hypersurface equation for bulk brane solutions. This was
\begin{equation}
uvw=\prod_{i=1}^{n-1}(t-f(q^ix))+\prod_{i=1}^{n-1}(t-g(q^iy))-t^n
\end{equation}
The singularities of this hypersurface equation are easy to locate, we search for points where all the partial derivatives vanish. The partial derivative with respect to $w$ shows that either $v$ or $u$ is $0$.  Assuming $v=0$ and the requirement that the original hypersurface equation is solved fixes $t$ to be equal to one of the $f(q^ix)$.  Therefore the partial derivative with respect to $t$ is satisfied only if $f(x)=f(q^ix)$ for some $i$ (i.e. there is a double root in $Q(t,u)$).  Likewise, if $u=0$ we get singularities at points where $g(y)=g(q^iy)$.  Note that unlike previous work where $U(Y)=0$, we do not have any remaining codimension two lines of singularities, these have been deformed leaving behind only codimension three singularities.

We now see that the fractional brane solutions directly correspond to the singularities found above.  We derived that fractional brane solutions exist at points where $f(x)=f(q^i x)$ or $g(y)=g(q^iy)$ for some $i$, which exactly correspond to the singularities of the hypersurface equation.
However, we have to be more careful. These are singularities of the Calabi-Yau geometry only 
if $x$ or $y$ are different from zero. This is because we have to take partial derivatives with respect to $u, v$ and not with respect to $x,y$ to see if the point of the Calabi-yau is singular or not. Away from the branching locus of $u=x^n$ and $v=y^n$ calculating by taking partial derivatives with respect to $x$ or $y$ is allowed. At the origin of the branching locus
one can actually calculate directly in the hypersurface equation and find that so long as
$a_1$ or $b_1$ are non-vanishing, then this is a fake singularity. Also in that case one can not solve for the fractional brane solutions as we did above, as both $X$ and $Y$ will be nilpotent, and neither can be argued to be diagonalizable.
Notice also that for each singularity there are two fractional branes associated with it.
If $f(x) = f(q^ix)$ for $i<n$, then $f((q^i x))= f(q^{n-i} (q^i x))$, and both of these solutions 
are associated to a brane sitting at the same value of $u$. A similar argument holds for $v$.
Also, one can check that for generic values of the superpotential coefficients, all of these singularities lead to conifolds (double points).

\begin{subsubsection}{Relevant Quantum Deformations of the Hypersurface}
We know that when considering the full quantum system these singularities could be resolved by quantum deformations of the 
hypersurface equation \cite{KS,GV,CV}. These are related to partial gaugino condensates \cite{DV}. We will show that the number of moduli of these deformations is exactly equal to the number
of fractional brane solutions divided by two (i.e. the number of singularities). The idea is that for each of these conifolds singularities we have one relevant quantum deformation (we will also be considering marginally relevant) as defined in \cite{CV}:  Given a Calabi-Yau manifold and its holomorphic 3-form $\Omega$, a deformation of the form $b_if(u,v,w,t)$ is relevant iff
\begin{equation}
\lim_{\Lambda\to \infty}\der{}{b_i}\oint_{B_j} \Omega\sim O(\log \Lambda)
\end{equation}
where $B_j$ are the set of non-compact cycles and $\Lambda$ is a cutoff required by the non-compactness of the $B_j$.  We can analyze what forms of deformation satisfy this by assigning charges and looking at the scaling behavior of the holomorphic 3-form.
\par
First, we assign charges to the variables so that ($F(u,v,t,w)=0$), the hypersurface equation, has charge $1$.  This fixes the degrees of the variables to be
\begin{align}
r_t&=\frac{1}{n}\\
r_u&=\frac{1}{l}\\
r_v&=\frac{1}{m}\\
r_w&=1-\frac{1}{l}-\frac{1}{m}
\end{align}
where $l$ is the degree of $f(x)$ and $m$ is the degree of $g(y)$.  The holomorphic 3-form is
\begin{equation}
\Omega=\frac{du\wedge dv\wedge dw}{\der{F}{t}}
\end{equation}
and has weight $r_u+r_v+r_w+r_t-1=\frac{1}{n}$.  If we have a deformation\footnote{This is the most general form since terms with both $u$ and $v$ can be absorbed into $w$.} of the form $\widetilde{F}=b_iu^at^b$, then 
the condition to be relevant is that
\begin{equation}
\left.\der{}{b_i}\frac{du\wedge dv\wedge dw}{\der{F+\widetilde F}{t}}\right|_{b_i=0}
\end{equation}
scales as less then 0 (this includes the marginal case). Taylor expanding and taking the only surviving term we get that
\begin{equation}
\der{\tilde F}{t}\frac{du\wedge dv\wedge dw}{\left(\der{F}{t}\right)^2}
\end{equation}
which scales as
\begin{equation}
\frac{a}{l}+\frac{b-1}{n}+\frac{1}{l}+\frac{1}{m}+1-\frac{1}{l}-\frac{1}{m}-2\left(1-\frac{1}{n}\right)
\end{equation}
Requiring that this be $<0$ gives
\begin{equation}
an+bl<l(n-1)
\end{equation}
The number of quantum moduli is therefore equal to the number of solutions $(a,b)$ to this equation for given integers $n,l,m$.  We also have quantum deformations of the form $\widetilde F=b_i v^at^b$, which are counted by the solutions
to
\begin{equation}
an+bm<m(n-1)
\end{equation}
\end{subsubsection}
\begin{subsubsection}{Counting Singularities}
Here we will show that the number of moduli is equal to the number of singularities or fractional brane solutions (actually pairs of fractional branes).
We found that there were singularities at points where $f(x)=f(q^ix)$ or $g(y)=g(q^iy)$.  Since $f$ is a degree $l$
polynomial in $x$ and $g$ is degree $m$, $f(x)-f(q^ix)$ is also a degree $l$ polynomial and has $l-1$ solutions, 
except in the case where $q^{il}=1$, where it has only $l-2$ solutions (we aren't counting $0$ as a solution, we'll add that  special case later).  Also, we note that there is a symmetry between 
the solutions, since if there is a root $x_0$ that satisfies $f(x)=f(q^ix)$ then $x_0'=q^ix_0$
satisfies $f(x)=f(q^{n-i}x)$.  Since $x_0^n={x'_0}^n$ these points are the same in the full  geometry associated to the $u,v$ variables.
Therefore we count the number of roots by
\begin{equation}
d(i)=\begin{cases}
l-2\quad li=0\mod n\\
l-1\quad{\rm otherwise}
\end{cases}
\end{equation}
and
\begin{equation}
e(i)=\begin{cases}
m-2\quad mi=0\mod n\\
m-1\quad{\rm otherwise}
\end{cases}
\end{equation}
as in \cite{HK}.  The total number of singularities is then
\begin{equation}
g=\begin{cases}
\sum_{i=1}^{(n-1)/2} d(i)+e(i)\quad n\;{\rm odd}\\
\frac{d(n/2)+e(n/2)}{2}+\sum_{i=1}^{(n-2)/2} d(i)+e(i)\quad n\;{\rm even}
\end{cases}
\end{equation}
plus $1$ for the singularity at $0$, which is actually made from $n$ identical fractional branes at that point.
Now the counting question has been reduced to a purely mathematical question:  Is $g$ equal to the number of solutions
of $an+bl<l(n-1)$ plus the number of solutions of $an+bm<m(n-1)$ (plus one to count $0$)?
The answer is yes, and although you can't perform the sum explicitly you can show that the sum counts the number of solutions of the inequalities exactly.  Therefore, for each singularity there exists a pair of fractional branes that behave like the Klebanov-Strassler setup \cite{KS},  and a gaugino condensate on these fractional branes resolve the associated singularity.  This way we count one quantum moduli per conifold of the deformed Calabi-Yau.
\end{subsubsection}
\end{subsection}
\subsection{Explicit Solutions for small $n$}
\begin{subsubsection}{$n=2$ Case}
Restricting to this case for illustrative purposes, we see that $q=-1$ and the above solutions become
\begin{align}
X&=x\left(\begin{array}{cc}
1&0\\0&-1
\end{array}\right)\\
Y&=y\left(\begin{array}{cc}
0&1\\
1&0
\end{array}\right)\\
Z&=\left(\begin{array}{cc}
z_1'&z_1-z\\
z_1+z&-z_1'
\end{array}\right)
\end{align}
The center elements are:
\begin{align}
u&=X^2=x^2\\
v&=Y^2=y^2\\
w&=-\frac{1}{2}\tr(Z^2)=\det Z=z^2-z_1^2-(z_1')^2\\
t&=xyz
\end{align}
(In this case, $-\frac{1}{2}\tr Z^2=\det Z$, since $Z$ is traceless.)
Therefore
\begin{align}
uvw&=x^2y^2(z^2-z_1^2-(z_1')^2)\\
&=t^2-x^2y^2z_1^2-x^2y^2(z_1')^2\\
&=t^2-x^2y^2\left(\sum_{j=1\mod n}y^{-1}\frac{a_{j}x^{j-1}}{2}\right)^2-x^2y^2\left(\sum_{j=1\mod n}x^{-1}\frac{b_{j}y^{j-1}}{2}\right)^2\\
uvw&=t^2-\frac{u}{4}\left(\sum_{j=0}a_{1+2j}u^j\right)^2-\frac{v}{4}\left(\sum_{j=0}b_{1+2j}v^j\right)^2
\end{align}
This is the hypersurface equation.  We can see this is equal to the general form derived earlier by noting:
\begin{align}
uvw&=t^2+f(x)f(-x)+g(y)g(-y)\\
&=(t-f(x))(t-f(-x))+(t-g(y))(t-g(-y))-t^2
\end{align}
since $f(x)+f(-x)=0$ in this case.
\par
Where are the singularities of this equation?  First, the partial derivative with respect to $t$ gives that
$t=0$, and the derivative with respect to $w$ gives that $uv=0$, therefore either $u$ or $v$ must be $0$.
The partial derivative with respect to $u$ gives:
\begin{equation}
vw=\frac{1}{2}\left(\sum_{j=0} a_{1+2j}u^j\right)\left(\sum_{j=0}a_{1+2j}\left(j+\frac{1}{2}\right)u^j\right)\label{n2x1}
\end{equation}
The partial derivative with respect to $v$ gives similarly:
\begin{equation}
uw=\frac{1}{2}\left(\sum_{j=0} b_{1+2j}v^j\right)\left(\sum_{j=0}b_{1+2j}\left(j+\frac{1}{2}\right)v^j\right)\label{n2x2}
\end{equation}
In the $v=0$ case, the first equation is solved by the roots of the rhs polynomial, and the second simply fixes
$w$ to a constant.  If $u=0$ it's the other way around. So our singularities are at these points:
\begin{align}
t=0,v=0,u_i={\rm Roots}\left[\left(\sum_{j=0} a_{1+2j}u^j\right)\right],w_i=\frac{b_1^2}{4u_i}\\
t=0,u=0,v_i={\rm Roots}\left[\left(\sum_{j=0} b_{1+2j}v^j\right)\right],w_i=\frac{a_1^2}{4v_i}
\end{align}
Of course, looking closer at these polynomials, we see that (with generic values for the constants) the roots are exactly the roots of
$f(x)$ and $g(y)$ squared as defined above.  The other roots of \eqref{n2x1} and \eqref{n2x2} and not singularities since that solution does not
satisfy the original hypersurface equation.
\par
The fraction brane solutions are trivial since they are one-dimensional.  Solving the F-term equations:
\begin{equation}
xy=0
\end{equation}
implies either $x$ or $y$ is $0$.  Let $y=0$, then:
\begin{align}
2zx&=-b_1\\
z&=-\frac{b_1}{2x}
\end{align}
and
\begin{align}
\sum_{i=0}a_{1+2i}x^{2i}&=0
\end{align}
This is exactly the singular points where $v=0$ as derived from the algebraic geometry above.  The $x=0$ case is identical.
\end{subsubsection}
\begin{subsubsection}{$n=3$ Case}
The general solution to the equations is:
\begin{align}
X&=x\left(\begin{array}{ccc}
1&0&0\\
0&q&0\\
0&0&q^2
\end{array}\right)\\
Y&=y\left(\begin{array}{ccc}
0&0&1\\
1&0&0\\
0&1&0
\end{array}\right)\\
Z&=\left(\begin{array}{ccc}
z_1'&z_1+z_2q+zq^2&z_2'q\\
z_2'&z_1'q^2&z_1+zq+z_2q^2\\
z+z_1+z_2&z_2'q^2&z_1'q
\end{array}\right)
\end{align}
The elements of the center are $u=X^3=x^3$, $v=Y^3=y^3$, and
\begin{equation}
w=\frac{1}{3}\tr(Z^3)=\det Z=z^3+z_1^3+z_2^3+(z_1')^3+(z_2')^3-3z(z_1z_2+z_1'z_2')
\end{equation}
(Again, in this case the trace is equal to the determinant expression, however that is no longer
true in $n\geq 4$).  As usual, $t=xyz$ (we always arrange for it this way to simplify the hypersurface equation).
The hypersurface equation is:
\begin{align}
uvw&=t^3+x^3y^3(z_1^3+z_2^3+(z_1')^3+(z_2')^3-3z(z_1z_2+z_1'z_2')\\
&=t^3-u\left(\sum_{j=0}\frac{a_{1+3j}u^j}{1-q}\right)^3-u^2\left(\sum_{j=0}\frac{a_{2+3j}u^j}{1-q^2}\right)^3\\
&\quad\quad\quad-v\left(\sum_{j=0}\frac{b_{1+3j}v^j}{1-q}\right)^3-v^2\left(\sum_{j=0}\frac{b_{2+3j}v^j}{1-q^2}\right)^3\\
&\quad\quad\quad-3tu\left(\sum_{j=0}\frac{a_{1+3j} u^j}{1-q}\right)\left(\sum_{j=0}\frac{a_{2+3j}u^j}{1-q^2}\right)-3tv\left(\sum_{j=0}\frac{b_{1+3j} v^j}{1-q}\right)\left(\sum_{j=0}\frac{b_{2+3j}v^j}{1-q^2}\right)
\end{align}
This actually factors as noted in the general solution, and can be written in a simpler form as:
\begin{equation}
uvw=\prod_{i=1}^3(t-f(xq^i))+\prod_{i=1}^3(t-g(yq^i))-t^3
\end{equation}
(there are no cross terms in this case either), with $f$ and $g$ given in equations \eqref{deff} and \eqref{defg}.  
\par
Let's derive this using the determinant form instead.  We can write the matrix $t-f(X)-g(qY)$ out explicitly in terms
of $t$ and the functions $f$ and $g$:
\begin{align}
\det\left(
\begin{array}{ccc}
t-f(x)&-xyz_2'&-xyz_1'\\
-xyz_1'&t-f(qx)&-xyz_2'\\
-xyz_2'&-xyz_1'&t-f(q^2x)
\end{array}\right)
\end{align}
Writing out the terms we get:
\begin{equation}
\prod_{i=0}^{2} (t-f(q^ix))-uv(z_1')^3-uv(z_2')^3-(3t-f(x)-f(qx)-f(q^2x))x^2y^2z_1'z_2'
\end{equation}
First we note that $f(x)+f(qx)+f(q^2x)=0$.  It is also easy to prove that $uv(z_1')^3+uv(z_2')^3=g(y)g(qy)g(q^2y)$ from
$g(y)=xyz_1'+xyz_2', g(qy)=qxyz_1'+q^2xyz_2'$, and $g(q^2y)=q^2xyz_1'+qxyz_2'$.  Therefore it simplifies to
\begin{equation}
\prod_{i=0}^{2} (t-f(q^ix))-g(y)g(qy)g(q^2y)-3tx^2y^2z_1'z_2'.
\end{equation}
We can also show directly by using the above that $g(y)g(qy)+g(qy)g(q^2y)+g(q^2y)g(y)=-3x^2y^2z_1'z_2'$, so 
we've derived explicitly that:
\begin{equation}
uvw=\prod_{i=0}^{2} (t-f(q^ix))+\prod_{i=0}^{2}(t-g(q^i y))-t^3
\end{equation}
as claimed.
So as in the $U(Y)=0$ cases, there are singular points
where $f(x)=f(xq^p)$ for $p<n$, and in addition we also have singular points at $g(y)=g(yq^p)$ as well.
\end{subsubsection}
\end{section}
\section{Matrix Model}
We now use the correlation between this theory and a bosonic matrix model, following the procedure used in \cite{Bhol}, and the combinatorial solutions to multi-matrix models worked
in \cite{Staud}.  Studying this related matrix model should
allow us to solve for the quantum corrected moduli space.  Our calculation above found the moduli space for a probe brane in the geometry, and we will split
the eigenvalues of the matrices in the matrix model into sets that correspond to bulk branes, $X$ type fractional branes, and $Y$ type fraction brane eigenvalues. 
We use this gauge choice to integrate out cross term matrices and derive an effective potential for the branes in the bulk.  This is our quantum moduli space for probe
branes as derived from the matrix model, and the derivation parallels exactly the one in \cite{Bhol}.
\par
The associated matrix model has potential
\begin{equation}
W(X,Y,Z)=N\mu^{-1}[\tr(XYZ-qXZY)+\tr\;V(X)+\tr\; U(Y)]
\end{equation}
with $n$ correlated eigenvalues corresponding to a single brane in the bulk.
\subsection{Quantum Deformations of Potentials}
To compute the quantum corrected moduli space, we follow the procedure in \cite{Bhol}.  We divide the matrices as follows:
\begin{equation}
X=\left(\begin{array}{ccccc}
X_b&0&\cdots&0&X_{bf_y}\\
0&\lambda_1&\ddots&\ddots&0\\
\vdots&\ddots&\ddots&\ddots&\vdots\\
0&\ddots&\ddots&\lambda_{n_x}&0\\
X_{f_yb}&0&\cdots&0&X_{f_y}
\end{array}\right)\quad
Y=\left(\begin{array}{ccccc}
Y_b&Y_{bf_x}&0&\cdots&0\\
Y_{f_xb}&Y_{f_x}&0&\vdots&0\\
0&\cdots&\lambda_1'&\ddots&0\\
\vdots&\ddots&\ddots&\ddots&0\\
0&0&0&0&\lambda_{n_y}'
\end{array}\right)\quad
Z=\left(\begin{array}{ccc}
Z_b&Z_{bf_x}&Z_{bf_y}\\
Z_{f_xb}&Z_{f_x}&Z_{f_xf_y}\\
Z_{f_yb}&Z_{f_yf_x}&Z_{f_y}
\end{array}\right)
\end{equation}
with $X_b$ as the $n$ by $n$ matrix corresponding to the brane in the bulk, $n_x$ eigenvalues for fractional branes on the $X$ type singularities,
and $n_y$ eigenvalues for fractional brane on $Y$ type singularities.  The partial diagonalization of $X$ and $Y$ is a gauge choice, and therefore 
gives you a Vandermonde determinant which gives a contribution to the effective potential of the form 
\begin{equation}
2\sum_{i}\tr_b\log(X_b-\lambda_i)+\sum_{i\neq j}\log(\lambda_i-\lambda_j)+2\sum_{i}\tr_b(\log(Y_b-\lambda'_i)+\sum_{i\neq j}\log(\lambda'_i-\lambda'_j)
\end{equation}
when you raise it up into the exponential.
\par
When you substitute this into the potential, it is easy to show that you get twice the number of terms as in \cite{Bhol}, and that $X_{bf_y},X_{f_yb},Y_{bf_x},Y_{f_xb}$
and all the off diagonal $Z$'s appear quadratically and can be integrated out, giving more contributions to the effective potential
\begin{equation}
-\sum_i \tr_b(\log(X_b-q\lambda_i))-\sum_i\tr_b(\log(X_b-q^{-1}\lambda_i))
-\sum_i\tr_b(\log(Y_b-q\lambda_i'))-\sum_i\tr_b(\log(Y_b-q^{-1}\lambda_i'))
\end{equation}
where we are ommiting terms that don't include the 
bulk brane matrices.  So in total the effective potential for the bulk brane in this model is:
\begin{align}
W=N\mu^{-1}(\tr(&(X_bY_bZ_b-qX_bZ_bY_b)+V(X_b)+U(Y_b)\\
&-\frac{\mu}{N}\sum_i\left[2\log(X_b-\lambda_i)-\log(X_b-q\lambda_i)-\sum_i\log(X_b-q^{-1}\lambda_i)\right]\\
&\left.-\frac{\mu}{N}\sum_i\left[2\log(Y_b-\lambda_i')-\log(Y_b-q\lambda_i')-\sum_i\log(Y_b-q^{-1}\lambda_i')\right]\right)
\end{align}
\par
%Need argument here about extra terms from $V(X)$ and $U(Y)$ and why they don't modify anything.
\par
At this point we simply have two copies of the results of \cite{Bhol}, and can apply the results of that paper.  The quantum version of $f(x)$ is
the same as in that case:
\begin{equation}
\widetilde{f}(X)=f(X)+\sum_{j=1}^{n-1}(1-q^i)X^{n-j}R_j(u)
\end{equation}
where 
\begin{equation}
R_j(\gamma)=\frac{1}{N}\sum_{k=0}^\infty\frac{\expect{\lambda^{nk+j}}}{\gamma^{k+1}}
\end{equation}
are the generating functions of the moments of the eigenvalues of $X$.  Likewise we now also have the quantum version of $g(Y)$:
\begin{equation}
\widetilde{g}(Y)=g(Y)+\sum_{j=1}^{n-1}(1-q^j)Y^{n-j}R_j'(v)
\end{equation}
where $R'$ is defined with the $\lambda'$.  
\par 
\subsection{Quantum Relations}
The point of using the matrix model to derive the quantum moduli space is that we claim these are equal to the deformations allowed by holomorphy we derived for the hypersurface
equation.  In fact we derived that the hypersurface equation is modified to:
\begin{equation}
\widetilde{Q}(t,u,v)=t^n+t^{n-2}(O_1(u)+O'_1(v))+t^{n-3}(O_2(u)+O'_2(v))+\cdots
\end{equation}
where the $O$'s and $O'$'s are polynomials in $u$ and $v$ respectively\footnote{The terms proportional to $t^{n-1}$ can be eliminated by linear shifts in $t$, or can 
also be seen to vanish from the scaling argument given above.}.  From the definitions given above, we can derive relations like:
\begin{align}
O_1(u)&=\sum_{i<j}\widetilde{f}(q^i x)\widetilde{f}(q^j x)\\
-O_2(u)&=\sum_{i<j<k}\widetilde{f}(q^i x)\widetilde{f}(q^j x)\widetilde{f}(q^k x)
\end{align}
and so on for the rest of the moduli.  These are the quantum relations for the matrix model, and there are also the corresponding equations for $O_i'(v)$ as
well.  Since these are identical to the quantum relations in \cite{Bhol}, we can use the results that these equations give recursion relations for the
moments of the eigenvalues $\expect{{\lambda^k}}$.  For example,
\begin{equation}
\sum_i a_i\expect{\lambda^{i+ns}}=\frac{\sum_j(2-q^j-q^{-j})\expect{\lambda^{ns-j}}\expect{\lambda^j}}{2}\label{qr}
\end{equation}
comes from requiring that the coefficients of the non-polynomial terms of the RHS of the first quantum relation vanish, as they must since
$O_1(u)$ is polynomial in $u$.  The second and higher quantum relations give similar recursion relations between the moments.  The identical equations also hold with $a_i\to b_i$ and $\lambda\to \lambda'$. 
\subsection{Loop Equations}
The quantum relations that we derived above from relating the matrix model to the $q$ deformed field field theory should also be encoded directly in the
loop equations of the matrix model.  The loop equations have the universal form of
\begin{equation}
\sum \mu \expect{f_1}\expect{f_2}=\left\langle\der{W}{\phi^i}\delta\phi^i\right\rangle
\end{equation}
where the sum is over all monomials $f_1,f_2$ such that $f_1\phi^if_2=f(\phi)$.  Here we are using a simplified notation where:
\begin{equation}
\expect{f(\phi)}=\left\langle\frac{1}{N}\tr(f(\phi))\right\rangle
\end{equation}
\subsubsection{First Equation}
By taking the variations $\delta Z=X^m Z X^{k-m}$, $\delta Z=Y^m Z Y^{k-m}$, $\delta X=X^{k+1}$ and $\delta Y= Y^{k+1}$, the loop equations we get are
\begin{align}
\mu\expect{X^m}\expect{X^{k-m}}&=\expect{YX^mZX^{k-m+1}}-q\expect{YX^{m+1}ZX^{k-m}}\\
\mu\expect{Y^m}\expect{Y^{k-m}}&=\expect{XY^{m+1}Z Y^{k-m}}-q\expect{XY^mZY^{k-m+1}}\label{sle}\\
\mu\sum_{m=0}^k\expect{X^m}\expect{X^{k-m}}-\expect{V'(X)X^{k+1}}&=\expect{X^{k+1}YZ}-q\expect{X^{k+1}ZY}\\
\mu\sum_{m=0}^k\expect{Y^m}\expect{Y^{k-m}}-\expect{U'(Y)Y^{k+1}}&=\expect{Y^{k+1}ZX}-q\expect{Y^{k+1}XZ}\label{fle}
\end{align}
It seems we have $k+2$ equations for the $k+2$ variables $\expect{YX^mZX^{k-m+1}}$ and another $k+2$ equations for the $k+2$ variables
$\expect{XY^{m}ZY^{k-m+1}}$ which determines them in terms of the $\expect{X^m}$ and $\expect{Y^m}$ respectively.  However when $k=0\mod n$, these
equations are no longer linearly independent, so for the equations to even have a solution, we sum the first $k+1$ equations weighted by $q$:
\begin{align}
\sum_{m=0}^k\mu q^m\expect{X^m}\expect{X^{k-m}}&=\sum q^m\expect{YX^mZX^{k-m+1}}-q\expect{YX^{m+1}ZX^{k-m}})\\
&=\expect{YZX^{k-m+1}}-q^{k-1}\expect{YX^{k+1}Z}\\
&=\mu\sum_{m=0}^k\expect{X^m}\expect{X^{k-m}}-\expect{V'(X)X^{k+1}}
\end{align}
Since $\expect{X^m}\expect{X^{k-m}}$ is symmetric over the sum from $0$ to $k$, we get that:
\begin{equation}
\expect{V'(X)X^{k+1}}=\frac{1}{2}\sum \mu(2-q^m-q^{-m})\expect{X^m}\expect{X^{k-m}}
\end{equation}
which is exactly the quantum relation \eqref{qr} derived above.  Clearly equations \eqref{sle} and \eqref{fle} will give the $v$ quantum relation.
All the comments given in \cite{Bhol} about twisted versus untwisted one-point functions still apply.
\subsubsection{Higher Equations}
The second and higher quantum relations should be obtained in the matrix model loop equations from higher term variations like $\delta Z=X^m Z^2 Y X^{k+1-m}$.
This is essentially the same technique as used in \cite{Bhol}, however we no longer can set words with unequal numbers of $Y$'s and $Z$'s to zero, since 
we no longer have the $U(1)$ charge symmetry between them.  We will look at variations and use the same idea as above.  
\par
First, we take the variation $\delta Z=X^m Z^2 YX^{k+1-m}$:
\begin{equation}
\mu\expect{X^mZ}\expect{YX^{k+1-m}}+\mu\expect{X^m}\expect{ZYX^{k+1-m}}=\expect{X^m Z^2YX^{k+2-m}Y}-q\expect{X^{m+1}Z^2YX^{k+1-m}Y}\label{sele}
\end{equation}
This is again $k+2$ equations for $k+3$ unknowns (the lower terms have already been solved for in the first order equations in terms of the $\expect{X^m}$),  so we 
need another equation to solve the system.  Actually it will be convieniant to have two extra equations (with one more unknown), which come from
the variations $\delta X=ZYX^{k+2}$ and $\delta X=YX^{k+2}Z$:
\begin{align}
\mu\sum_{m=0}^{k+1}\expect{ZYX^m}\expect{X^{k+1-m}}-\expect{ZYX^{k+2}V'(X)}&=\expect{YZZYX^{k+2}}-q\expect{ZYZYX^{k+2}}\label{sele1}\\
\mu\sum_{m=0}^{k+1}\expect{YX^m}\expect{X^{k+1-m}Z}-\expect{YX^{k+2}ZV'(X)}&=\expect{ZYZYX^{k+2}}-q\expect{ZZYYX^{k+2}}\label{sele2}
\end{align}
Again in the case where $k=0\mod n$, these equations are linearly independent and we get a consistency condition by summing them.  By summing \eqref{sele}
you can rewrite in terms of the sum of \eqref{sele1} and \eqref{sele2}, and you get the constraint:
\begin{align}
\nonumber \sum_{m=0}^{k+1}(1-q^m)\mu\expect{X^m}\expect{ZYX^{k+1-m}}+\sum_{m=0}^{k+1}(q-q^m)\mu\expect{X^mZ}\expect{YX^{k+1-m}}=\\
\expect{ZYX^{k+2}V'(X)}+q\expect{YX^{k+2}ZV'(X)}
\end{align}
But we can see from the variation $\delta Z=X^k$ that $\expect{YX^k}=0$, so the equation simplifes to:
\begin{equation}
\sum_{m=0}^{k+1}(1-q^m)\mu\expect{X^m}\expect{ZYX^{k+1-m}}=\expect{ZYX^{k+2}V'(X)}+q\expect{YX^{k+2}ZV'(X)}
\end{equation}
Now this equation has reduced to essentially the same equation as in \cite{Bhol}.  You can solve
the $\expect{ZYX^k}$ terms from the first order equations in terms of the $\expect{X^k}$, and that determines a recursion relation between them.  
The modification of the above to derive the second order $Y$ quantum relation is clear.


\begin{thebibliography}{99}

%\cite{Seiberg:1994bp}
\bibitem{Shol}
N.~Seiberg,
``The Power of holomorphy: Exact results in 4-D SUSY field theories,''
arXiv:hep-th/9408013.
%%CITATION = HEP-TH 9408013;%%

%\cite{Klebanov:2000hb}
\bibitem{KS}
I.~R.~Klebanov and M.~J.~Strassler,
``Supergravity and a confining gauge theory:
 Duality cascades and  chiSB-resolution of naked singularities,''
JHEP {\bf 0008}, 052 (2000)
[arXiv:hep-th/0007191].
%%CITATION = HEP-TH 0007191;%%



%\cite{Gopakumar:1998ii}
\bibitem{GV}
R.~Gopakumar and C.~Vafa,
``M-theory and topological strings. I,''
arXiv:hep-th/9809187.
%%CITATION = HEP-TH 9809187;%%

%\cite{Cachazo:2002pr}
\bibitem{CV}
F.~Cachazo and C.~Vafa,
``N = 1 and N = 2 geometry from fluxes,''
arXiv:hep-th/0206017.
%%CITATION = HEP-TH 0206017;%%
F.~Cachazo, B.~Fiol, K.~A.~Intriligator, S.~Katz and C.~Vafa,
``A geometric unification of dualities,''
Nucl.\ Phys.\ B {\bf 628}, 3 (2002)
[arXiv:hep-th/0110028].
%%CITATION = HEP-TH 0110028;%%
F.~Cachazo, S.~Katz and C.~Vafa,
``Geometric transitions and N = 1 quiver theories,''
arXiv:hep-th/0108120.
%%CITATION = HEP-TH 0108120;%%
F.~Cachazo, K.~A.~Intriligator and C.~Vafa,
``A large N duality via a geometric transition,''
Nucl.\ Phys.\ B {\bf 603}, 3 (2001)
[arXiv:hep-th/0103067].
%%CITATION = HEP-TH 0103067;%%


%\cite{Benvenuti:2004wx}
\bibitem{BHK}
  S.~Benvenuti, A.~Hanany and P.~Kazakopoulos,
  ``The toric phases of the Y(p,q) quivers,''
  JHEP {\bf 0507}, 021 (2005)
  [arXiv:hep-th/0412279].
 %%CITATION = HEP-TH 0412279;%%
  S.~Franco, A.~Hanany, D.~Martelli, J.~Sparks, D.~Vegh and B.~Wecht,
  ``Gauge theories from toric geometry and brane tilings,''
  arXiv:hep-th/0505211.
%%CITATION = HEP-TH 0505211;%%
 


 %\cite{Berenstein:2001jr}
\bibitem{BL}
D.~Berenstein and R.~G.~Leigh,
``Resolution of stringy singularities by non-commutative algebras,''
JHEP {\bf 0106}, 030 (2001)
[arXiv:hep-th/0105229].
%%CITATION = HEP-TH 0105229;%%

%\cite{Berenstein:2002ge}
\bibitem{Brev}
D.~Berenstein,
``Reverse geometric engineering of singularities,''
JHEP {\bf 0204}, 052 (2002)
[arXiv:hep-th/0201093].
%%CITATION = HEP-TH 0201093;%%



%\cite{Dijkgraaf:2002fc}
\bibitem{DV}
R.~Dijkgraaf and C.~Vafa,
``Matrix models, topological strings, and supersymmetric gauge theories,''
arXiv:hep-th/0206255.
%%CITATION = HEP-TH 0206255;%%

%\cite{DV2}
\bibitem{DV2}
R.~Dijkgraaf and C.~Vafa,
``On geometry and matrix models,''
arXiv:hep-th/0207106.
%%CITATION = HEP-TH 0207106;%%

%\cite{DV3}
\bibitem{DV3}
R.~Dijkgraaf and C.~Vafa,
``A perturbative window into non-perturbative physics,''
arXiv:hep-th/0208048.
%%CITATION = HEP-TH 0208048;%%

%\cite{Dijkgraaf:2002xd}
\bibitem{DGLVZ}
R.~Dijkgraaf, M.~T.~Grisaru, C.~S.~Lam, C.~Vafa and D.~Zanon,
``Perturbative computation of glueball superpotentials,''
arXiv:hep-th/0211017.
%%CITATION = HEP-TH 0211017;%%

%\cite{Cachazo:2002ry}
\bibitem{CDSW}
F.~Cachazo, M.~R.~Douglas, N.~Seiberg and E.~Witten,
``Chiral rings and anomalies in supersymmetric gauge theory,''
JHEP {\bf 0212}, 071 (2002)
[arXiv:hep-th/0211170].
%%CITATION = HEP-TH 0211170;%%


%\cite{Seiberg:1994rs}
\bibitem{SW}
  N.~Seiberg and E.~Witten,
  ``Electric - magnetic duality, monopole condensation, and confinement in N=2
  supersymmetric Yang-Mills theory,''
  Nucl.\ Phys.\ B {\bf 426}, 19 (1994)
  [Erratum-ibid.\ B {\bf 430}, 485 (1994)]
  [arXiv:hep-th/9407087].
  %%CITATION = HEP-TH 9407087;%%
H.~Itoyama and A.~Morozov,
``The Dijkgraaf-Vafa prepotential in the context of general  Seiberg-Witten theory,''
arXiv:hep-th/0211245.
%%CITATION = HEP-TH 0211245;%%


%\cite{Berenstein:2002sn}
\bibitem{Bq}
D.~Berenstein,
``Quantum moduli spaces from matrix models,''
Phys.\ Lett.\ B {\bf 552}, 255 (2003)
[arXiv:hep-th/0210183].
%%CITATION = HEP-TH 0210183;%%


%\cite{Berenstein:2003cz}
\bibitem{Bhol}
  D.~Berenstein,
``Solving matrix models using holomorphy,''
  JHEP {\bf 0306}, 019 (2003)
  [arXiv:hep-th/0303033].
%%CITATION = HEP-TH 0303033;%%

%\cite{Ferrari:2003vp}
\bibitem{Fer}
  F.~Ferrari,
  ``Planar diagrams and Calabi-Yau spaces,''
  Adv.\ Theor.\ Math.\ Phys.\  {\bf 7}, 619 (2004)
  [arXiv:hep-th/0309151].
  %%CITATION = HEP-TH 0309151;%%
%\cite{Bonelli:2005dc}
\bibitem{Bon}
  G.~Bonelli, L.~Bonora and A.~Ricco,
  ``Conifold geometries, topological strings and multi-matrix models,''
  Phys.\ Rev.\ D {\bf 72}, 086001 (2005)
  [arXiv:hep-th/0507224].
  %%CITATION = HEP-TH 0507224;%%

%\cite{Berenstein:2001uv}
\bibitem{Bcon}
D.~Berenstein,
``On the universality class of the conifold,''
JHEP {\bf 0111}, 060 (2001)
[arXiv:hep-th/0110184].
%%CITATION = HEP-TH 0110184;%%

%\cite{Dymarsky:2005xt}
\bibitem{DKS}
  A.~Dymarsky, I.~R.~Klebanov and N.~Seiberg,
  ``On the moduli space of the cascading SU(M+p) x SU(p) gauge theory,''
  JHEP {\bf 0601}, 155 (2006)
  [arXiv:hep-th/0511254].
  %%CITATION = HEP-TH 0511254;%%


%\cite{Berenstein:2005xa}
\bibitem{BHOP}
  D.~Berenstein, C.~P.~Herzog, P.~Ouyang and S.~Pinansky,
  ``Supersymmetry breaking from a Calabi-Yau singularity,''
  JHEP {\bf 0509}, 084 (2005)
  [arXiv:hep-th/0505029].
  %%CITATION = HEP-TH 0505029;%%
  \bibitem{dP1}
  S.~Franco, A.~Hanany, F.~Saad and A.~M.~Uranga,
  ``Fractional branes and dynamical supersymmetry breaking,''
  arXiv:hep-th/0505040.
  %%CITATION = HEP-TH 0505040;%%
  M.~Bertolini, F.~Bigazzi and A.~L.~Cotrone,
  ``Supersymmetry breaking at the end of a cascade of Seiberg dualities,''
  Phys.\ Rev.\ D {\bf 72}, 061902 (2005)
  [arXiv:hep-th/0505055].
  %%CITATION = HEP-TH 0505055;%%

%\cite{Intriligator:2005aw}
\bibitem{IS}
  K.~Intriligator and N.~Seiberg,
``The runaway quiver,''
  arXiv:hep-th/0512347.
  %%CITATION = HEP-TH 0512347;%%


\bibitem{ME}
S.~Pinansky,
``Quantum Deformations from Toric Geometry,''
arXiv:hep-th/0511027.
%%CITATION = HEP-TH 0511027


%\cite{Douglas:1998xa}
\bibitem{D}
M.~R.~Douglas,
``D-branes and discrete torsion,''
arXiv:hep-th/9807235.
%%CITATION = HEP-TH 9807235;%%

%\cite{Berenstein:2000hy}
\bibitem{BL2}
  D.~Berenstein and R.~G.~Leigh,
  ``Discrete torsion, AdS/CFT and duality,''
  JHEP {\bf 0001}, 038 (2000)
  [arXiv:hep-th/0001055].
  %%CITATION = HEP-TH 0001055;%%


%\cite{Leigh:1995ep}
\bibitem{LS}
R.~G.~Leigh and M.~J.~Strassler,
``Exactly marginal operators and duality in four-dimensional N=1 supersymmetric gauge theory,''
Nucl.\ Phys.\ B {\bf 447}, 95 (1995)
[arXiv:hep-th/9503121].
%%CITATION = HEP-TH 9503121;%%

\bibitem{BJL}
D.~Berenstein, V.~Jejjala and R.~G.~Leigh,
``Marginal and relevant deformations of N = 4 field theories and
non-commutative moduli spaces of vacua,''
Nucl.\ Phys.\ B {\bf 589}, 196 (2000)
[arXiv:hep-th/0005087].
%%CITATION = HEP-TH 0005087;%%


%\cite{Dorey:2002pq}
\bibitem{DHK}
N.~Dorey, T.~J.~Hollowood and S.~P.~Kumar,
``S-duality of the Leigh-Strassler deformation via matrix models,''
JHEP {\bf 0212}, 003 (2002)
[arXiv:hep-th/0210239].
%%CITATION = HEP-TH 0210239;%%
T.~J.~Hollowood,
``New results from glueball superpotentials and matrix models: The
Leigh-Strassler deformation,''
arXiv:hep-th/0212065.
%%CITATION = HEP-TH 0212065;%%
  T.~Mansson,
  ``Another Leigh-Strassler deformation through the matrix model,''
  JHEP {\bf 0303}, 055 (2003)
  [arXiv:hep-th/0302077].
  %%CITATION = HEP-TH 0302077;%%


%\cite{Berenstein:2004ys}
\bibitem{BC}
  D.~Berenstein and S.~A.~Cherkis,
  ``Deformations of N = 4 SYM and integrable spin chain models,''
  Nucl.\ Phys.\ B {\bf 702}, 49 (2004)
  [arXiv:hep-th/0405215].
  %%CITATION = HEP-TH 0405215;%%


%\cite{Lunin:2005jy}
\bibitem{LM}
  O.~Lunin and J.~Maldacena,
  ``Deforming field theories with U(1) x U(1) global symmetry and their gravity
  duals,''
  JHEP {\bf 0505}, 033 (2005)
  [arXiv:hep-th/0502086].
  %%CITATION = HEP-TH 0502086;%%

%\cite{Maldacena:1997re}
\bibitem{M}
  J.~M.~Maldacena,
  ``The large N limit of superconformal field theories and supergravity,''
  Adv.\ Theor.\ Math.\ Phys.\  {\bf 2}, 231 (1998)
  [Int.\ J.\ Theor.\ Phys.\  {\bf 38}, 1113 (1999)]
  [arXiv:hep-th/9711200].
  %%CITATION = HEP-TH 9711200;%%


%\cite{Gukov:1998kk}
\bibitem{Guk}
  S.~Gukov,
  ``Comments on N = 2 AdS orbifolds,''
  Phys.\ Lett.\ B {\bf 439}, 23 (1998)
  [arXiv:hep-th/9806180].
  %%CITATION = HEP-TH 9806180;%%

%\cite{Berenstein:2003sr}
\bibitem{BSoo}
  D.~Berenstein and S.~J.~Rey,
  ``Wilsonian proof for renormalizability of N = 1/2 supersymmetric field
  theories,''
  Phys.\ Rev.\ D {\bf 68}, 121701 (2003)
  [arXiv:hep-th/0308049].
  %%CITATION = HEP-TH 0308049;%

%\cite{Douglas:1999hq}
\bibitem{DF}
M.~R.~Douglas and B.~Fiol,
``D-branes and discrete torsion. II,''
arXiv:hep-th/9903031.

%\cite{Berenstein:2003fx}
\bibitem{Brun}
  D.~Berenstein,
  ``D-brane realizations of runaway behavior and moduli stabilization,''
  arXiv:hep-th/0303230.
  %%CITATION = HEP-TH 0303230;%%


\bibitem{Detal}
N.~Dorey, ``A new deconstruction of little string theory,'' 
JHEP {\bf 0407}, 016 (2004)
[arXiv:hep-th/0406104].
%%CITATION = HEP-TH 0406104
N.~Dorey, ``S-duality, deconstruction and confinement for a marginal deformation of N=4 SUSY Yang-Mills,''
JHEP {\bf 0408}, 043 (2004)
[arXiv:hep-th/0310117].
%%CITATION = HEP-TH 0310117
N,~Dorey, T.~J.~Hollowood,
``On the Coulomb Branch of a Marginal Deformation of ${\cal N}=4$ SUSY Yang-Mills,''
JHEP {0506}, 036 (2005)
[arXiv:hep-th/0411163].
%%CITATION = HEP-TH 0411163

%\cite{Aharony:2002hx}
\bibitem{Ket}
  O.~Aharony, B.~Kol and S.~Yankielowicz,
  ``On exactly marginal deformations of N = 4 SYM and type IIB  supergravity on
  AdS(5) x S**5,''
  JHEP {\bf 0206}, 039 (2002)
  [arXiv:hep-th/0205090].
  %%CITATION = HEP-TH 0205090;%%
  B.~Kol,
  ``On conformal deformations,''
  JHEP {\bf 0209}, 046 (2002)
  [arXiv:hep-th/0205141].
  %%CITATION = HEP-TH 0205141;%%

%\cite{Polchinski:2000uf}
\bibitem{PS}
  J.~Polchinski and M.~J.~Strassler,
  ``The string dual of a confining four-dimensional gauge theory,''
  arXiv:hep-th/0003136.
  %%CITATION = HEP-TH 0003136;%%


%\cite{Hollowood:2004dc}
\bibitem{HK}
T.~J.~Hollowood and S.~P.~ Kumar,
``An ${\cal N}=1$ duality cascade from a deformation of ${\cal N}=4$ SUSY Yang-Mills theory,''
arXiv:hep-th/0407029.
%%CITATION = HEP-TH 0407029%%

%\cite{Staudacher:1993xy}
\bibitem{Staud}
M.~Staudacher,
``Combinatorial solution of the two matrix model,''
Phys.\ Lett.\ B {\bf 305} (1993) 332
[arXiv:hep-th/9301038].
%%CITATION = HEP-TH 9301038;%%








\end{thebibliography}
\end{document}